\documentstyle[aps,manuscript]{revtex}

\begin{document}

\title {Crowd-anticrowd theory of the Minority Game}
\author{M. Hart, P. Jefferies and N.F. Johnson}
\address {Physics Department, Oxford University,
Oxford, OX1 3PU, U.K.}
\author{P.M. Hui}
\address {Department of Physics, The Chinese
University of Hong Kong, Shatin, \\
New Territories, Hong Kong}

\maketitle

\begin{abstract}

The  
Minority Game is a simple yet highly non-trivial agent-based model for a 
complex adaptive system. Despite its importance, a quantitative explanation of
the game's fluctuations which applies over the entire parameter range of
interest has so far been lacking.  We provide such a quantitative
description based on the interplay between crowds of
like-minded agents and their anti-correlated partners (anticrowds).

\end{abstract}
\bigskip


\newpage

Agent-based
models of complex adaptive systems are attracting significant interest across a
variety of disciplines\cite{arthur}. An important application currently receiving
much attention within the physics community, is the study of fluctuations
in financial markets\cite{econophysics}. Each agent knows the past ups and downs
in the price index of a financial market and must decide how to trade based on
this global information. These decisions then feed back to generate the
fluctuations or `volatility' of the price index itself. The resulting dynamical,
many-body system is hence highly non-linear; indeed the combination of
adaptability and frustration make such agent-based systems
arguably richer and more challenging than those typically studied in
condensed matter and statistical physics. 

The Minority Game (MG) introduced by
Challet and Zhang\cite{challet,savit}, offers possibly the simplest paradigm for
such a complex, adaptive system. The development of a complete, quantitative
theory of the MG is of fundamental importance - the MG has therefore been the
subject of intense research activity in the physics
community\cite{econophysics,challet,savit,spinglass,sherrington,dHR,cavagna,us,crowd}.
The most striking feature arising from numerical simulations\cite{savit} is the
non-monotonic variation in the `volatility' $\sigma$ (i.e. the standard deviation
of the fluctuations) of the `price' as the agents'
memory-length $m$ is varied. Challet {\em et al} provided a
sophisticated formal connection between the MG and spin glass
systems\cite{spinglass} yielding many fascinating quantitative
results for the MG's dynamics. However, no general theory has yet been
provided which yields quantitative agreement with the numerical results for
$\sigma$
\cite{savit} over the entire range of $m$, $s$ and $N$ values. 

This paper shows that a theoretical
model can be constructed in a surprisingly simple way by incorporating
the `crowd' effects (i.e. strong inter-agent correlations) which arise
within the
interacting, many-agent population. The results yield very good agreement
with the numerical results for $\sigma$  
over the entire range of $m$, $s$ and $N$.
The
non-monotonic behaviour of the volatility $\sigma$ \cite{challet,savit} 
results from a fascinating interplay between the actions of groups of
like-minded agents (`crowds') and the actions of opposing groups
(`anticrowds').

The MG\cite{challet} comprises an
odd number of agents $N$ who choose repeatedly between
option 0 (e.g. buy) and option 1 (e.g. sell).   
The winners are those in the
minority group, e.g. sellers win if there is an excess of buyers. The
outcome at
each timestep represents the winning decision, 0 or 1. 
A common bit-string of the $m$ most recent
outcomes is made available to the agents at each timestep.
The agents
randomly pick $s$ strategies at the beginning of the game, with repetitions
allowed. After each turn, the agent assigns one (virtual) point to each of his
strategies which would have
predicted the correct outcome. At each turn of the game, the agent
uses the most successful strategy, i.e. the one with the 
most virtual points, 
among his $s$ strategies. The strategy-space ${\cal V}_m$ forms a
$2^m$-dimensional  hypercube with
strategies at the $2^{2^m}$ vertices. Fortunately,
the MG's standard deviation is essentially unchanged if a `reduced' strategy
space ${\cal U}_m$ is used instead of ${\cal V}_m$ \cite{challet}: the
${\cal U}_m$ only contains
$2^{m+1}$ strategies or equivalently $2^m$ strategy pairs $\{\cal G\}$.
The two strategies within a given pair $\cal G$ are anticorrelated, i.e. they
differ by the maximum possible Hamming distance $d_H=2^m$ \cite{challet}.
Strategies between any two pairs
$\cal G$ and $\cal G'$ are uncorrelated, i.e. they differ by $d_H=2^{m-1}$.  The
results presented in this paper employ the reduced strategy space. 

If ${n}_R$ agents use the same strategy $R$, then they
will act as a `crowd', i.e. they will make the same decision. If
${n}_{\bar R}$ agents simultaneously use the 
anticorrelated strategy  
$\bar R$, they will make the opposite decision and will hence act as an
`anticrowd' (${\cal G}\equiv (R,\bar R)$). If
${n}_R\approx {n}_{\bar R}$ for all $\cal G$, then the actions of the crowds and
anticrowds cancel and the standard deviation $\sigma$ of the number of agents
$A(t)$ choosing a given option  will be small. In contrast if
${n}_R
\gg n_{\bar R}$ for all
$\cal G$, then
$\sigma$ will be large. Since there is no correlation between $\cal G$ and
$\cal
G'$, each group $\cal G$ comprising a
crowd-anticrowd pair $(n_R,n_{\bar R})$ can be taken as contributing to $A(t)$
via a separate, essentially random walk in time of step-size
$|n_R-n_{\bar R}|$. The variances of these walks can be summed to
obtain the standard deviation of $A(t)$. Hence the MG 
standard deviation is given by
\begin{equation}
\sigma = \bigg[\sum_{{\cal G}=1}^{2^{m}} \sigma^2_{\cal
G}\bigg]^{\frac{1}{2}} =
\bigg[\sum_{{\cal G}\equiv (R,{\bar R})}
\frac{1}{4}|n_R-n_{\bar R}|^2\bigg]^{\frac{1}{2}}
\
\end{equation}
where both time-averaging, for a given configuration of
initial strategies, and configuration-averaging have been
carried out. We now demonstrate that this crowd-anticrowd cancellation underlies
the numerical results for
$\sigma$ vs. $m$
\cite{savit}.
We run the numerical simulation of the MG and wait until transients in
$A(t)$
have disappeared. At timestep $t_0$, we read out the number of players
playing each strategy $R$, where $R=1,2,\dots 2^{m+1}$. For each strategy pair
${\cal G}=(R,{\bar R})$, we obtain
$n_R-n_{\bar R}$ at time $t_0$ and hence calculate $\sigma$.  We then
average this $\sigma$ over 1000 timesteps to simulate the time-averaging (our
results are insensitive to the precise time-averaging procedure).  Finally,
we
average over 16 runs to simulate the configuration-averaging.
Figure 1 compares the resulting time and configuration-averaged $\sigma$ with that
obtained from the numerical MG simulation. The agreement is very good for all
$m$, $s$ and
$N$ (not shown). Hence the crowd-anticrowd cancellation quantitatively explains
the numerical results for $\sigma$ \cite{savit}.

Since the labels $R$ are arbitrary in Eq. (1), the ordering of
strategies
$\{n_R\}$ has no particular significance.
At any particular time
$t_0$, these
$2^{m+1}$ strategies can be ranked according to their virtual points by a
sort-operation $\Theta$ acting on the list $\{n_R\}$. Hence
$\{n_R\}\stackrel{\Theta}\mapsto \{{n}_\rho\}$ where $\rho$ is
the virtual-point rank label with $\rho=1$ being the highest scoring
strategy and
$\{n_\rho\}\equiv n_{\rho=1}, n_{\rho=2},\dots n_{\rho={2^{m+1}}}$.
Alternatively, strategies can be ranked according to their popularity $\{n_r\}$ 
where
$\{n_r\}\equiv n_{r=1}, n_{r=2},\dots n_{r={2^{m+1}}}$, i.e.
strategy $r=1$ is that used by the largest number of agents,
strategy $r=2$ is that used by the second largest number of agents, etc. 
Note that $\{n_r\}$ can be
obtained from $\{n_R\}$ and hence $\{n_\rho\}$ by sort operations, i.e.
$\{n_R\}\stackrel{\Psi}\mapsto \{n_r\}$ and
hence $\{n_\rho\}\stackrel{\Gamma}\mapsto
\{n_r\}$.
Since each agent plays
the available strategy having highest virtual points, it follows that
\begin{equation} n_r=N\bigg(\bigg[1 - \frac{(r-1)}{2^{m+1}}\bigg]^s -
\bigg[1-\frac{r}{2^{m+1}}\bigg]^s\bigg)
\end{equation}
where $\sum_{r=1}^{2^{m+1}} n_r =N$ as required. Agents are discrete
objects; hence the simulation tends to produce discrete steps in the curves of
$n_r$ as a function of $r$, particularly  for large
$m$ where the total number of strategies $2^{m+1}$
exceeds $N$. Hence we convert the theoretical
$n_{r}$
values of Eq. (2) to an integer. For large $m$ such that $2^{m+1}\gg N$, the
resulting theoretical values are typically $n_r\sim 1$ for small $r$ and
$n_r=0$ for $r>N$. We have checked that the theoretical and numerical values for 
$n_{r}$ agree for general $m$. 

For the
virtual-point ordered list $\{n_\rho\}$, the strategy
$\rho'=2^{m+1}+1-\rho$ is {\em always} anticorrelated to the strategy
$\rho$, i.e.
$\rho'\equiv \bar \rho$. Hence knowledge of the sort operation $\Gamma$
completely
determines where  each strategy's anticorrelated partner is located in the
popularity-ordered list $\{n_r\}$. Since we are here only interested
in time-averaged and run-averaged
$\sigma$, we only need to consider the probability distribution of locations of
$\bar r$ in the  popularity-ordered
list $\{n_r\}$. We
therefore replace the sort operation $\Gamma$ by a
probability function $P(r'={\bar r})$ which gives the
probability that any strategy $r'$ is the anti-correlated partner
of strategy $r$ in the list $\{n_r\}$.
Hence Eq. (1) becomes
\begin{equation}
\sigma =
\bigg[ \frac{1}{2} \sum_{r=1}^{2^{m+1}} \sum_{r'=1}^{2^{m+1}}
\frac{1}{4}|n_r-n_{r'}|^2 P(r'={\bar r})\bigg]^{\frac{1}{2}}
\
\ \ .
\end{equation}
When the virtual-point ordered
list
$\{n_\rho\}$ and the popularity-ordered list $\{n_r\}$ are identical, then
$P(r'={\bar r})$ will be a $\delta$-function at $r'=2^{m+1}+1-r$ and hence
Eq. (3)
has the same form 
as Eq. (1). In the opposite case where the two ordered
lists
are uncorrelated, $P(r'={\bar r})$ should be a flat distribution.

Figure 2 shows $P(r'={\bar r})$ for $r=1$ as a function of $r'$, taken from
the
numerical MG  simulation at $m=2,5$ and $10$. For small $m$ ($m=2$) the
anticorrelated strategy to the most popular strategy (i.e. $r=1$) 
is at $r'=2^{m+1}$,
i.e. it is the least popular strategy. Hence $P(r'={\bar r})$ resembles the
$\delta$-function limiting case mentioned above. 
Very few agents will therefore play this anticorrelated strategy. Hence the
crowd-anticrowd
cancellation will be small and $\sigma$ will be large, as can be seen in
Fig. 1. As
$m$ increases ($m=5$) a remarkable effect occurs: the peak in $P(r'={\bar
r})$ moves
up toward $r=1$. Hence both $r=1$ and its anticorrelated partner $\bar r$
are now very
popular. For $m=2$ it seemed like there was an effective `repulsion'
between
$r$ and $\bar r$; for $m=5$ this seems more like an attraction yielding
a bound `exciton-like' pair with the crowd (anticrowd) playing the role of the
positive (negative) charge. For large $m$ ($m=10$), the ability of the
anticrowd to `screen' the crowd has decreased yielding a rather flat
distribution. The strong crowd-anticrowd correlation which
appears as $m$
increases means that the crowd and anticrowd become comparable in size. Hence
$\sigma$ is small for $m\sim 5-6$, in agreement with Fig. 1. 
Note that the MG cannot fully `optimize' itself by building equal-sized crowds and
anticrowds because of the strategy-space frustration built in at the beginning
of the game.  In modified MG models such as Ref. \cite{prl} 
which features evolving stochastic
strategies,  this frustration is
able to relax - similar sized crowds
and anti-crowds hence emerge yielding a smaller-than-random
$\sigma$ for general $m$.

For small $m$, 
the two lists $\{n_\rho\}$ and $\{n_r\}$
will be very similar, hence $P(r'={\bar r}) \sim \delta_{r',2^{m+1}+1-r}$. The
discreteness of the agents is now unimportant since $n_r\gg1$,
hence $n_r$ can
be treated as continuous. Equations (2) and (3) yield
\begin{eqnarray}
\sigma_{{\rm low}\ m} & = & \frac{N}{2}\bigg[ \sum_{r=1}^{2^{m}} \bigg[
\bigg(1-\frac{r-1}{2^{m+1}}\bigg)^s - \bigg(1-\frac{r}{2^{m+1}}\bigg)^s
\nonumber \\
& & - \bigg(\frac{r}{2^{m+1}}\bigg)^s + \bigg(\frac{r}{2^{m+1}}\bigg)^s
\bigg(1 -
\frac{1}{r}\bigg)^s
\bigg]^2\
\bigg]^{\frac{1}{2}}\ \ .
\end{eqnarray}
For $s=2$ this becomes
\begin{equation}
\sigma_{{\rm low}\ m} = \frac{N}{{\sqrt 3}\  2^{\frac{m}{2}+1}} \bigg[ 1 -
2^{-2(m+1)}\bigg] ^{\frac{1}{2}} \ \ .
\end{equation}
The inset in Fig. 3 shows these analytic curves for $s=2$ and $s=4$ (thick solid
lines monotonically decreasing). As
might be expected using the extreme
$\delta$-function form for
$P(r'={\bar r})$, these curves produce {\em upper} estimates of
$\sigma$ for small $m$. Now consider the opposite extreme of uncorrelated $r'$
and
$\bar r$,
i.e. the flat distribution
$P(r'={\bar r})\sim 2^{-(m+1)}$.
For $s=2$ this gives
\begin{equation}
\sigma_{{\rm low}\ m} = \frac{N}{{\sqrt 3}\  2^{\frac{(m+3)}{2}}} \bigg[ 1 -
2^{-2(m+1)}\bigg] ^{\frac{1}{2}} \ \ .
\end{equation}
Equation (6), and its generalization for $s>2$, produce
{\em lower} estimates (thin dashed lines in Fig. 3 inset). 
For a given $s$, the $\sigma$ values 
obtained from separate numerical runs tend to be scattered between the two
corresponding limiting analytic curves. For larger $m$ ($m>6$) the
discreteness of the agents implies that
$n_r\sim 1$ for $r<N$ while
$n_r=0$ for
$r>N$. Using the integer form of Eq. (2) for high $m$, and the flat distribution
for $P(r'={\bar r})$, yields
\begin{equation}
\sigma_{{\rm high}\ m} =  \frac{\sqrt N}{2} \bigg[ 1 -
\frac{N}{2^{m+1}}\bigg]^{\frac{1}{2}}\ .
\end{equation}
Hence 
$\sigma$ approaches the coin-toss limit from below as $m\rightarrow
\infty$ (thick solid line monotonically increasing in Fig. 3 inset) as observed
numerically. The analytic curves in Fig. 3
(inset) have further similarities with the numerical results of Fig.
1:  (a) a minimum in
$\sigma$ for $s=2$, (b) minimum moves to higher
$m$ as
$s$ increases, (c) minimum becomes shallower as $s$ increases, (d) $\sigma$ is
not sensitive to $s$ for large $m$. Equation (7) also shows
that
$\sigma_{{\rm high}\ m}\rightarrow 0$ at $N=2^{m+1}$, i.e. for $m\sim 5-6$.
Using conditional probabilities to go one step beyond the approximation of a
flat $P(r'={\bar r})$ distribution, 
we find that $\sigma_{{\rm high}\ m}$ for general $s$ is given by 
${\sqrt N}[1 -
2^{-(m+1)} (Ns-1)]^{\frac{1}{2}}/2$. Interestingly, this form is consistent with
the results of the spin-glass replica approach\cite{challetdisc,us3}.

Figure 3 (main graph) shows the
spread of numerical values for different runs compared to
the full crowd-anticrowd theoretical calculation. The
appropriate analytic expressions for the probability
function
$P(r'={\bar r})$ in Eq. (3) involve multiple sums and are cumbersome:
we therefore obtained the results for each $m$ in Fig. 3 by generating the
corresponding 
$P(r'={\bar r})$ similar to those in Fig. 2. The agreement is very good.
The theoretical points tend to lie toward the high end of the
numerical spread, for example at
$m=2$; this can be attributed to the fact that our theory neglects accidental
degeneracies in the virtual-point ordered list $\{n_\rho\}$. We have checked that
including a stochastic (i.e. coin-toss) process to break such ties, reduces the
theoretical
$\sigma$ values down toward the mid-point of the numerical spread hence making
the agreement even better\cite{us3}.
 
In summary we have presented an analytic analysis of crowding
effects in the MG which provides quantitative agreement with the main
finding of Ref.
\cite{savit} across the entire parameter
range of interest.

We thank D. Challet for his comments.

\begin{figure}
\bigskip

\caption{Standard deviation $\sigma$ for the Minority Game as a function of
memory-size $m$
for $s=2,3,4$ strategies per agent and $N=101$ agents. Solid curve:
numerical
simulation. Dashed curve: crowd-anticrowd theory using Eq.
(1). Random (coin-toss) limit $\sigma={\sqrt N}/2=5.0$ is indicated.}
\bigskip
\bigskip

\caption{Probability function $P(r'={\bar r})$ giving the probability that the
strategy
ranked $r'$ on the popularity-ordered list, is anti-correlated with the
strategy
ranked $r$. Results are shown for $r=1$ (i.e. most popular strategy) as a
function of
$r'$ for
$m=2$ (dotted-dashed), $m=5$ (dotted) and
$m=10$ (solid).
$s=2$ and $N=101$. Note that $\sum_{r'} P(r'={\bar r}) =1$.}
\bigskip
\bigskip

\caption{Theoretical crowd-anticrowd calculation (solid circles) vs. numerical 
simulations (open circles)
for
$s=2$, $N=101$.
16 numerical runs are shown for each $m$. Inset: 
Theoretical curves for $\sigma$ using limiting analytic forms in Eqs.
(4)-(7).}

\end{figure}

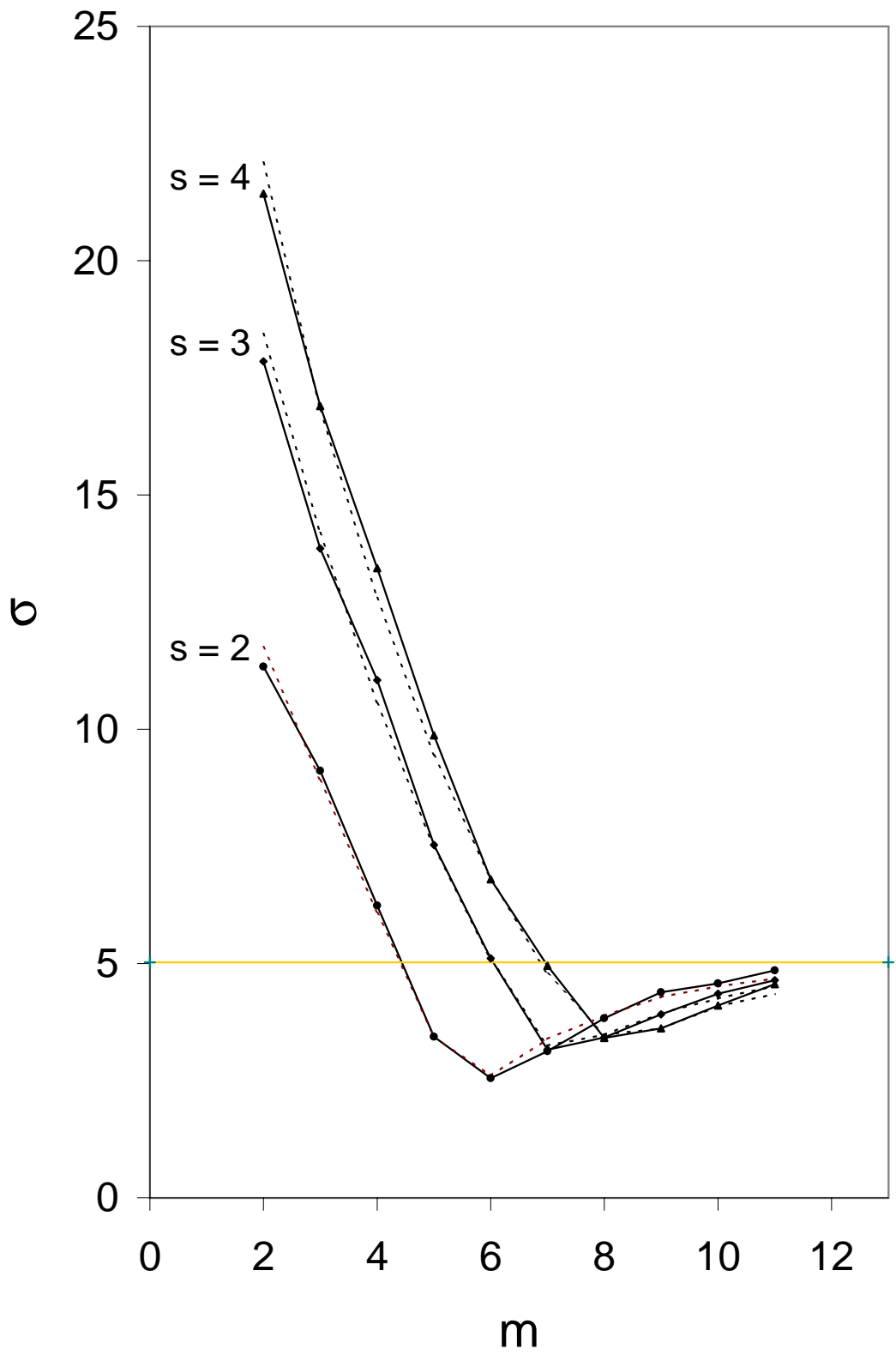
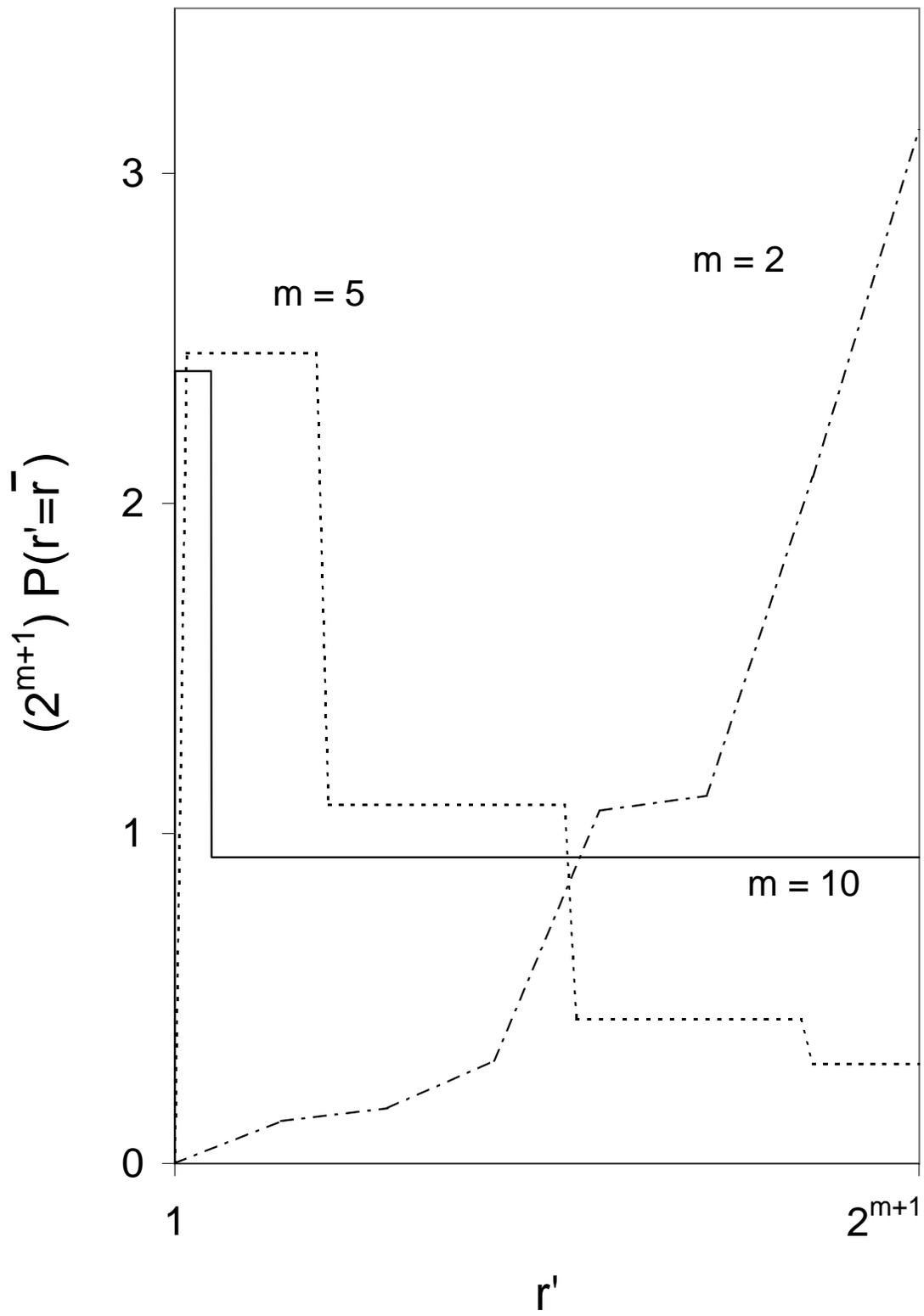
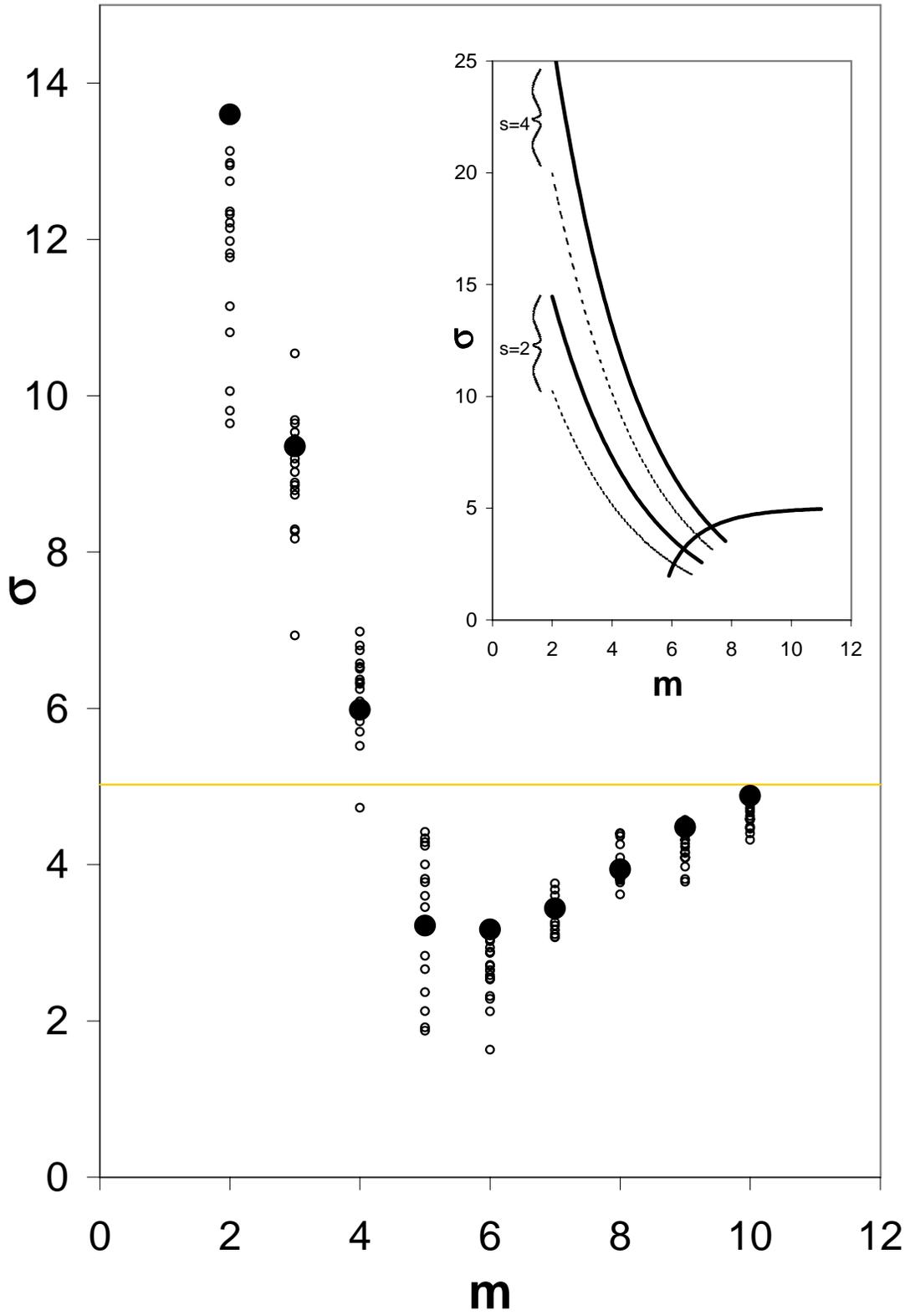
\end{document}